\documentclass[preprint,aps,prd,showpacs,nofootinbib]{revtex4}
\usepackage{graphicx}
\usepackage{amsmath}
\usepackage{amssymb}
\usepackage{bm}
\usepackage{bbm}
\begin{document}
\title{
Invariant-mass distribution of $\bm{c}\bar{\bm{c}}$
in $\bm{\Upsilon(1S)\to}\bm{c}\bar{\bm{c}}+\bm{X}$}
%%%%%%%%%%%%%%%%%%%%%%%%%%%%%%%%%%%%%%%%%%%%%%%%%%%%%%%%%%%%%%%%%%%%%%%%%%%%%%
\author{Hee~Sok~Chung}
\affiliation{Department of Physics, Korea University, Seoul 136-701, Korea}
\author{Taewon~Kim}
\affiliation{Department of Physics, Korea University, Seoul 136-701, Korea}
\author{Jungil~Lee}
\affiliation{Department of Physics, Korea University, Seoul 136-701, Korea}
\date{\today}
%%%%%%%%%%%%%%%%%%%%%%%%%%%%%%%%%%%%%%%%%%%%%%%%%%%%%%%%%%%%%%%%%%%%%%%%%%%%%%
\begin{abstract}
We calculate the invariant-mass distribution for the $c\bar{c}$ pair
produced in the inclusive $\Upsilon(1S)$ decay based on the
color-singlet mechanism of the nonrelativistic quantum chromodynamics
factorization approach at leading order in the bottom-quark velocity
$v_b$ in the meson rest frame. As the short-distance processes, we
consider $b\bar{b}\to g^*gg$ followed by $g^*\to c\bar{c}$ and
$b\bar{b}\to \gamma^*\to c\bar{c}$ at leading order in the strong
coupling. The invariant-mass distribution of the $b\bar{b}\to c\bar{c}gg$
contribution has a sharp peak just above the threshold and that of
the $b\bar{b}\to\gamma^*\to c\bar{c}$ channel is concentrated at the 
maximally allowed kinematic end point. We predict that
$\Gamma[\Upsilon(1S)\to c\bar{c}+X]/%
\Gamma[\Upsilon(1S)\to \textrm{light hadrons}]= 0.065\,\alpha_s$,
which is smaller than a previous result by about $20\,\%$.
\end{abstract}
% 12.38.-t   Quantum chromodynamics
% 12.39.St  Factorization
% 13.20.Gd  Decays of J/psi, Upsilon, and other quarkonia
% 14.40.Gx   Mesons with S=C=B=0, mass > 2.5 GeV (including quarkonia)
%\keywords{}
\maketitle
\section{\label{sec:intro}Introduction}
%--------------------------------------------------------------------
Recent experimental analyses being carried out by the CLEO III
experiment on the inclusive charm production in bottomonia
decays~\cite{CLEO} have activated a series of theoretical 
studies~\cite{Bodwin:2007zf,Kang:2007uv,Hao:2007rb}
based on the nonrelativistic quantum chromodynamics (NRQCD)
factorization approach~\cite{Bodwin:1992ye,Bodwin:1994jh}.
Predictions for the branching fractions and the charmed-hadron 
momentum distributions in inclusive $\chi_{bJ}$ ($J=0$, $1$, and $2$) 
decays were presented in Ref.~\cite{Bodwin:2007zf}.
These processes are nice probes to investigate the color-octet 
mechanism of the NRQCD factorization approach which is distinguished 
from the color-singlet-model calculation in 
Ref.~\cite{Barbieri:1979gg}. 
In addition to the spin-triplet $P$-wave (${}^3P_J$) bottomonium
decay, the CLEO Collaboration also analyzes the charm production in
the spin-triplet $S$-wave (${}^3S_1$) bottomonium decay.
Early theoretical studies on the inclusive charm production in
the inclusive $\Upsilon(1S)$ decay began in the late 1970s. In 1978,
Fritzsch and Streng calculated the invariant-mass distribution
of the $c\bar{c}$ pair produced in the inclusive $\Upsilon(1S)$
decay~\cite{Fritzsch:1978ey} by considering the QCD process
$b\bar{b}\to g^*gg\to c\bar{c}gg$ within the color-singlet model,
where they predicted the branching fraction
$\textrm{Br}[\Upsilon(1S)\to c\bar{c}+X]$ to be a few percents
in the limit that the charm-quark momentum $\bm{p}_c^*$ in
the $c\bar{c}$ rest frame can be neglected. In 1996,
Cheung, Keung, and Yuan calculated the $J/\psi$ production
rate with the same short-distance process, where they
considered the decay of the color-singlet $b\bar{b}$ pair
producing a $c\bar{c}$ pair in the color-octet spin-triplet
state that evolves into the $J/\psi$~\cite{Cheung:1996mh}.
One can find more studies on the bottomonium decay in
Refs.~\cite{Mackenzie:1981sf,Keung:1982jb,%
Campbell:2007ws}.

In this paper, based on the color-singlet mechanism of the
NRQCD factorization approach, we compute the invariant-mass
distribution of the $c\bar{c}$ pair 
in $\Upsilon(1S)\to c\bar{c}+X$. This is an extension of a
recent work on the total charm production rate and the momentum
distribution of the charm hadrons produced in the inclusive 
$\Upsilon(nS)$ decay~\cite{Kang:2007uv}.
As is studied in Ref.~\cite{Kang:2007uv}, we consider
the decay of the color-singlet spin-triplet $b\bar{b}$ pair
[$b\bar{b}_1({}^3S_1)$] into $g^*gg$ followed
by $g^*\to c\bar{c}$, which we call the QCD contribution.
We also consider the short-distance process
$b\bar{b}_1({}^3S_1)\to\gamma^*\to c\bar{c}$, which we call
the QED contribution. The perturbative calculations of the
short-distance processes are carried
out at leading order in the strong coupling $\alpha_s$ and
the QED coupling $\alpha$. For the long-distance part of the
NRQCD factorization formula, we consider the leading contribution
with respect to the bottom-quark velocity $v_b$ in the $\Upsilon(1S)$
rest frame. The relevant NRQCD matrix element for the channel is
$\langle\Upsilon(1S)| \mathcal{O}_1({}^3S_1)|\Upsilon(1S)\rangle$
defined in Ref.~\cite{Bodwin:1994jh}, where $\mathcal{O}_1({}^3S_1)$
is the color-singlet spin-triplet four-quark operator for the 
annihilation decay of the $\Upsilon(1S)$.
At higher orders in $v_b$, the color-octet short-distance processes
$b\bar{b}_8({}^3S_1)\to g^*\to c\bar{c}$, $b\bar{b}_8({}^1S_0)\to
g^*g\to c\bar{c}g$, and $b\bar{b}_8({}^3P_J)\to g^*g\to c\bar{c}g$
can also contribute to the charm production in the inclusive
$\Upsilon(1S)$ decay. The color-octet contributions are estimated
to be about $10\,\%$ of the color-singlet contributions
$b\bar{b}_1({}^3S_1)\to c\bar{c}gg$ and
$b\bar{b}_1({}^3S_1)\to\gamma^*\to c\bar{c}$~\cite{Kang:2007uv}.
Therefore, we neglect the color-octet processes in this work.

This paper is organized as follows: In Sec.~\ref{sec:cc}, we present
the NRQCD factorization formula for the inclusive charm production
in $\Upsilon(1S)$ decay and compute the short-distance coefficients 
for the $c\bar{c}$ invariant-mass distribution of the process.
Numerical analysis for the invariant-mass distribution is given
in Sec.~\ref{sec:rate}, which is followed by a summary 
in Sec.~\ref{summary}.
%--------------------------------------------------------------------
\section{\label{sec:cc}Charm-quark production in $\bm{\Upsilon(1S)}$ 
decay}
%--------------------------------------------------------------------
In this section, we present the NRQCD factorization formula for the
$c\bar{c}$ invariant-mass distribution in $\Upsilon(1S)\to c\bar{c}+X$
within the color-singlet mechanism at leading order in $v_b$. As the 
short-distance contributions, we consider the QCD process 
$b\bar{b}_1({}^3S_1)\to g^*gg$ followed by $g^*\to c\bar{c}$ and 
the QED process $b\bar{b}_1({}^3S_1)\to \gamma^*\to c\bar{c}$.
We follow the formalism to calculate the inclusive charm production 
rate in the $\Upsilon(nS)$ decay in Ref.~\cite{Kang:2007uv}.
%--------------------------------------------------------------------
\subsection{\label{sec:NRQCD}NRQCD factorization formula}
%--------------------------------------------------------------------
Within the color-singlet mechanism of the NRQCD factorization
formalism the differential rate for the inclusive decay
$\Upsilon(1S)\to c\bar{c}+X$ at leading order in $v_b$ is given
by~\cite{Kang:2007uv}
%------------------
\begin{equation}
\label{Gam-c}%
%------------------
d\Gamma[\Upsilon(1S) \to c\bar{c} + X] =
dC_1^{(c)}\,
\frac{\langle \mathcal{O}_1({}^3S_1) \rangle_{\Upsilon(1S)}}{m_b^2},
%------------------
\end{equation}
%------------------
where $\langle \mathcal{O}_1({}^3S_1) \rangle_{\Upsilon(1S)}=%
\langle\Upsilon(1S)| \mathcal{O}_1({}^3S_1)|\Upsilon(1S)\rangle$
is the leading-order color-singlet NRQCD matrix element for the
$\Upsilon(1S)$ and $m_b$ is the mass of the bottom quark. The 
short-distance coefficient $dC_1^{(c)}$ is insensitive to the
long-distance nature of the $\Upsilon(1S)$ and calculable as a 
perturbative series of the strong coupling $\alpha_s$. At leading
order in $\alpha_s$ and $\alpha$, the dimensionless short-distance
coefficient $dC_1^{(c)}$ is the sum of the QCD and QED contributions:
%------------------
\begin{equation}
%------------------
\label{dc1}%
dC_1^{(c)} =
dC_1^{(\textrm{QCD})}+dC_1^{(\textrm{QED})}.
%------------------
\end{equation}
%------------------
As is mentioned earlier, the QCD process denotes
$b\bar{b}_1({}^3S_1)\to c\bar{c}gg$ followed by $g^*\to c\bar{c}$
and the QED process stands for 
$b\bar{b}_1({}^3S_1)\to\gamma^*\to c\bar{c}$. We use the identifiers
(QCD) and (QED) to denote those contributions, respectively, 
in the remainder of this paper.

The short-distance coefficients can be determined by 
perturbative matching. If we replace the heavy-quarkonium state
$|\Upsilon(1S)\rangle$ in Eq.~(\ref{Gam-c}) with the perturbative
$b\bar{b}_1({}^3S_1)$ state, $|b\bar{b}_1({}^{3}S_1)\rangle$, then
the long-distance matrix element $\langle
\mathcal{O}_1({}^{3}S_1)\rangle_{\Upsilon(1S)}$ is replaced by
the perturbative NRQCD matrix element $\langle \mathcal{O}_1({}^{3}S_1)%
\rangle_{b\bar{b}_1({}^{3}S_1)}$, while the short-distance
coefficient $dC_1^{(c)}$ remains the same:
%------------------
\begin{equation}
\label{pert-fac}%
%------------------
d\Gamma[b\bar{b}_1({}^3S_1)\to c\bar{c}+X]
=dC_1^{(c)}\frac{
\langle \mathcal{O}_1({}^{3}S_1)\rangle_{b\bar{b}_1({}^{3}S_1)}
}{m_b^2}.
%------------------
\end{equation}
%------------------
Both the left side of Eq.~(\ref{pert-fac}) and the matrix element
$\langle \mathcal{O}_1({}^{3}S_1)\rangle_{b\bar{b}_1({}^{3}S_1)}$
are calculable perturbatively.
By taking the ratio of the two quantities, we can determine the
short-distance coefficient $dC_1^{(c)}$.
%------------------------------------------------
\subsection{\label{sec:amp-qcd}Short-distance coefficients }
%------------------------------------------------
In this section, we describe the procedure to compute the
short-distance coefficient $dC_1^{(\textrm{QCD})}$ for the QCD process.
We also provide the expression for $dC_1^{(\textrm{QED})}$ by quoting a
previous result in Ref.~\cite{Kang:2007uv}. In both short-distance
processes, the momenta of the $b$ and the $\bar{b}$ can be expressed
in terms of the total momentum $P$ and the relative momentum $q$ of
the $b\bar{b}$ pair:
%------------------
\begin{subequations}
\begin{eqnarray}
%------------------
p&=&\tfrac{1}{2}P+q,
\\
\bar{p}&=&\tfrac{1}{2}P-q,
%------------------
\end{eqnarray}
\end{subequations}
%------------------
where $p$ and $\bar{p}$ satisfy the on-shell conditions
$p^2=\bar{p}^2=m_b^2$ and $P\cdot q=0$. In the rest frame of the
$b\bar{b}$ pair, $P=(2E_b,0)$ and $q=(0,\bm{q})$, where
$E_b=\sqrt{m_b^2+\bm{q}^2}$.

At leading order in $\alpha_s$, the full QCD
amplitude for the short-distance process $b(p)\bar{b}(\bar{p})\to
c(p')\bar{c}(\bar{p}') g(p_2) g(p_3)$ is given by
%------------------
\begin{eqnarray}
\label{amp-qcd}%
%------------------
\mathcal{M}_{b\bar{b}}^{(\textrm{QCD})} &=&
\frac{(4\pi \alpha_s)^2}{(p'+\bar{p}')^2}
  \bar{u}(p') T^a \gamma_\lambda v(\bar{p}')
  \epsilon^{b *}_{2\,\sigma}(p_2)\epsilon^{c *}_{3\,\tau}(p_3)
\nonumber
\\
&&
\times
\sum_\textrm{perm} \bar{v}(\bar{p})
\left[ \gamma^\lambda \frac{1}{\not\!p-\not\!p_2-\not\!p_3-m_b}
\gamma^\sigma \frac{1}{\not\!p-\not\!p_3-m_b}\gamma^\tau
\otimes T^a T^b T^c \right] u(p),
%------------------
\end{eqnarray}
%------------------
where $\sum_\textrm{perm}$ denotes the summation over the
permutations of the three gluons attached to the bottom-quark line,
$T^a$ is a generator of the SU(3)-color in the
fundamental representation, $a$, $b$, and $c$ are color indices for
the gluons, and $\epsilon_2$ and $\epsilon_3$ are polarization
four-vectors of the external gluons with momenta $p_2$ and $p_3$,
respectively.

In order to extract the $b{\bar b}_1({}^3S_1)$ contribution from the
$b\bar{b}$ amplitude (\ref{amp-qcd}), we employ the
covariant-projection method~\cite{Kuhn:1979bb,Guberina:1980dc,%
Bodwin:2002hg}, which replaces the outer product
$u(p) {\bar v}({\bar p})$ of the $b\bar{b}$ spinors with
the direct product of the color-singlet projector $\pi_1$
and the spin-triplet projector $\epsilon\cdot \Pi_3$,
which are defined by
%------------------
\begin{subequations}
\label{projectors}%
\begin{eqnarray}
%------------------
\label{color-projector}%
\pi_1 &=& \frac{1}{\sqrt{N_c}} \mathbbm{1},
\\
\label{spin-projector}%
\epsilon\cdot\Pi_3 &=& -\frac{1}{4\sqrt{2}E_b(E_b+m_b)}
(/\!\!\!{p}+m_b)(\,/\!\!\!\!P\!+\!2E_b) \,/\!\!\!\epsilon\,
(/\!\!\!\bar{p}-m_b),
%------------------
\end{eqnarray}
\end{subequations}
%------------------
where $N_c=3$ is the number of colors,
$\mathbbm{1}$ is the unit matrix of the SU(3)-color,
and $\epsilon$ is the polarization four-vector of the
$b\bar{b}_1({}^3S_1)$ state so that $P \cdot \epsilon  = 0$. The
projectors (\ref{projectors}) are normalized as
$\textrm{Tr}[\pi_1\pi_1^\dagger]=1$ and
$\textrm{Tr}[(\epsilon\cdot\Pi_3)%
(\epsilon\cdot\Pi_3)^\dagger]=4p_0\bar{p}_0$.

Because we are to compute the $c\bar{c}$ invariant-mass
distribution, it is convenient to write the amplitude (\ref{amp-qcd})
as the product of the vector current $\mathcal{C}^\mu$ for 
$g^*\to c\bar{c}$ and the amputated amplitude $\mathcal{B}^{\mu}$
for $b\bar{b}_1({}^3S_1)\to ggg^*$, where
%------------------
\begin{subequations}
\label{current}%
\begin{eqnarray}
%------------------
\mathcal{B}^\mu&=&
\sum_\textrm{perm} \textrm{Tr}\left[
\gamma^\mu\frac{1}{\not\!p-\not\!p_2-\not\!p_3-m_b}\not\!\epsilon_2
\frac{1}{\not\!p-\not\!p_3-m_b}\not\!\epsilon_3 \epsilon\cdot\Pi_3
\right],\\
\mathcal{C}^\mu&=&
\frac{1}{(p'+\bar{p}')^2} \bar{u}(p') \gamma^\mu v(\bar{p}').
%------------------
\end{eqnarray}
\end{subequations}
%------------------
Note that we do not include the color factor and the coupling
in Eq.~(\ref{current}). At leading order in $v_b$, the
amplitude for the QCD process $b\bar{b}_1({}^3S_1)\to c\bar{c} gg$
becomes
%------------------
\begin{equation}
\label{amp-S}%
%------------------
\mathcal{M}_{b\bar{b}_1({}^3S_1)}^{(\textrm{QCD})} =
\frac{
(4\pi\alpha_s)^2
}{4\sqrt{N_c}}
d^{abc}T^a_{ij}
\mathcal{B}_\mu \mathcal{C}^\mu
\biggl\vert_{q=0},
%------------------
\end{equation}
%------------------
where $i$ and $j$ are color indices for the charm quark and the
charm antiquark, respectively. In Eq.~(\ref{amp-S}), we put $q=0$ 
to take the $v_b$-leading contributions so that $E_b=m_b$, used
Tr$[T^aT^bT^c]=(d^{abc}+if^{abc})/4$, and chose only the symmetric
component of the color factor. At $v_b=0$, which we take at leading
order in $v_b$, the amplitude (\ref{amp-S}) is infrared (IR) finite
in the soft limits of any external gluons. However, in the limit that
the charm quark becomes massless, the amplitude (\ref{amp-S}) may
acquire collinear divergences.  We will return to this point later
in Sec.~\ref{sec:collinear}.

Squaring the amplitude (\ref{amp-S}) for
$b\bar{b}_1({}^3S_1)\to c\bar{c} gg$, averaging over the spin-triplet
states, and summing over the spins of the final states, we obtain the
differential annihilation rate for the QCD process.
The contribution of the QCD process $b\bar{b}_1({}^3S_1)\to c\bar{c}gg$
to the left side of Eq.~(\ref{pert-fac}) is
%------------------
\begin{equation}
\label{dgam-qcd}%
d\Gamma^{(\textrm{QCD})}_{b\bar{b}_1({}^3S_1)}=
\frac{(N_c^2-1)(N_c^2-4)}{N_c^2}
\frac{8\pi^4\alpha_s^4}{3}\,
\mathcal{B}^{\mu\nu}\mathcal{C}_{\mu\nu}
\frac{d\Phi_4}{2!},
\end{equation}
%------------------
where $d\Phi_4$ is the phase space element for the $c\bar{c}gg$
final state and the factors 1/3 and $1/2!$
are for the average over the initial spins and for the two
identical particles (gluons) in the  final state, respectively.
The tensors $\mathcal{B}^{\mu\nu}$ and $\mathcal{C}^{\mu\nu}$
in Eq.~(\ref{dgam-qcd}) are defined by
%------------------
\begin{subequations}
\label{tensor}%
\begin{eqnarray}
\mathcal{B}^{\mu\nu}&=&
\sum_{\textrm{spins}}\mathcal{B}^\mu \mathcal{B}^{* \nu},
\\
\mathcal{C}^{\mu\nu}&=&\sum_{\textrm{spins}}
\mathcal{C}^\mu \mathcal{C}^{* \nu}
\nonumber \\
&=&
\frac{
\textrm{Tr}\left[
(/\!\!\!{p'}+m_c)\gamma^\mu (/\!\!\!{\bar{p}'}-m_c)\gamma^\nu \right]
}{(p'+\bar{p}')^4}.
\end{eqnarray}
\end{subequations}
%------------------
The polarizations for the external gluons and the 
$b\bar{b}_1({}^3S_1)$ pair in $\mathcal{B}^{\mu\nu}$ are summed as
%------------------
\begin{subequations}
\label{Imunu}%
\begin{eqnarray}
%------------------
\sum_\lambda \epsilon_i^\alpha(\lambda)\epsilon_i^{*\beta}(\lambda)
&=&-g^{\alpha\beta}~\textrm{ for }~i=2,\textrm{ }3,
\\
\sum_\lambda \epsilon^\alpha(\lambda)\epsilon^{*\beta}(\lambda)
&=&-g^{\alpha\beta}+\frac{P^\alpha P^\beta}{P^2}.
%------------------
\end{eqnarray}
\end{subequations}
%------------------
By substituting Eq.~(\ref{dgam-qcd}) to the left side of
Eq.~(\ref{pert-fac}) and using the following value for
the perturbative NRQCD matrix element
%------------------
\begin{equation}
\label{me-pert-norm}%
\langle\mathcal{O}_1({}^3S_1)
\rangle_{b\bar{b}_1({}^3S_1)}
=2N_c (2E_b)^2= 8N_cm_b^2+\mathcal{O}(v_b^2),
\end{equation}
%------------------
we determine the short-distance coefficient $dC_1^{(\textrm{QCD})}$ as
%------------------
\begin{equation}
\label{dC}%
dC_1^{(\textrm{QCD})}=
\frac{(N_c^2-1)(N_c^2-4)}{N_c^3}
\frac{\pi^4\alpha_s^4}{3}
\mathcal{B}^{\mu\nu}\mathcal{C}_{\mu\nu}
\frac{d\Phi_4}{2!}.
\end{equation}
%------------------

For the QED process $bb_1({}^3S_1)\to \gamma^*\to c\bar{c}$,
we quote the result given in Ref.~\cite{Kang:2007uv}, which was
obtained by making use of the short-distance coefficient for
the leptonic decay of $\Upsilon(1S)$:
%------------------
\begin{equation}
\label{dc1-qed}%
%------------------
dC_1^{(\textrm{QED})}
=
\frac{\pi}{3}e_b^2e_c^2N_c\alpha^2 (2+r)
\sqrt{1-r}\,\delta(1-\xi)d\xi,
%------------------
\end{equation}
%------------------
where $e_Q$ is the fractional electric charge of the heavy quark
for $Q=c$, $b$ and the dimensionless variables $r$ and $\xi$
are defined by
%------------------
\begin{subequations}
\label{variables1}
\begin{eqnarray}
\label{variables1a}
   r&=&\frac{m_c^2}{m_b^2},\\
\label{variables1b}
\xi &=& \frac{m_{c\bar{c}}^2}{P^2},
\end{eqnarray}
\end{subequations}
%------------------
where $m_{c\bar{c}}$ is the invariant mass of the $c\bar{c}$ pair
and $m_c$ is the charm-quark mass.
The factor $\sqrt{1-r}$ in Eq.~(\ref{dc1-qed}) is the ratio of
the phase space for the $c\bar{c}$ final state to the massless
two-body phase space.
%------------------------------------------------
\subsection{\label{sec:phase-space}Phase-space integral
for the QCD process}
%------------------------------------------------
In order to compute the $c\bar{c}$ invariant-mass distribution of
the QCD process, it is convenient to factor out the two-body phase
space of the $c\bar{c}$ pair $d\Phi_2(p_1\to p'+\bar{p}')$
from the four-body phase space $d\Phi_4$ in Eq.~(\ref{dC}), where
$p_1$, $p'$, and $\bar{p}'$ are the momenta for the $c\bar{c}$
pair, the $c$, and the $\bar{c}$, respectively. Then $d\Phi_4$
becomes the product of $d\Phi_2(p_1\to p'+\bar{p}')$ and
the three-body phase space $d\Phi_3(P\to p_1+p_2+p_3)$,
where $p_2$ and $p_3$ are the momenta for the external gluons,
convolved with the $c\bar{c}$ invariant mass $m_{c\bar{c}}$ as
%-------------------
\begin{equation}
\label{dPhi4_1}
d\Phi_4(P\to p'+\bar{p}' +p_2+p_3)
=d\Phi_3(P\to p_1+p_2+p_3)\frac{dm_{c\bar{c}}^2}{2\pi}d\Phi_2(p_1\to p'+\bar{p}').
\end{equation}
%-------------------
The phase space (\ref{dPhi4_1}) can further be
simplified by using the dimensionless variable $\xi$ in 
Eq.~(\ref{variables1b}) and the scaled energy fraction $x_i$
in the $b\bar{b}$ rest frame:
%-------------------
\begin{equation}
\label{x_i}
x_i = \frac{2P\cdot p_i}{P^2}
\end{equation}
%-------------------
for $i=1$, 2, and 3. Rewriting the phase space (\ref{dPhi4_1}) in 
terms of $\xi$ and $x_i$, we get
%-------------------
\begin{equation}
\label{dPhi4_2}
d\Phi_4
=\frac{dx_1 dx_2 dx_3}{128\pi^3}\delta(2-x_1-x_2-x_3)
\frac{P^4 d\xi}{2\pi} \frac{|\bm{p}_c^*|d\Omega^*}{4m_{c\bar{c}}(2\pi)^2},
\end{equation}
%-------------------
where $\Omega^*$ is the solid angle of the charm quark with
the three momentum $\bm{p}_c^*$ in the $c\bar{c}$ rest frame.
The scaled energy fraction $x_3$ can be integrated out by using the
energy delta function. Then the $c\bar{c}$ invariant-mass
distribution is obtained as a double integral of $x_1$
and $x_2$, where the physical ranges for $\xi$, $x_1$, and $x_2$
are given by
%-------------------
\begin{subequations}
\label{ranges}
\begin{eqnarray}
\label{ranges-r}
r \leq &\xi& \leq 1, \\
2\sqrt{\xi} \leq &x_1& \leq 1+\xi, \\
\frac{1}{2}\bigg( 2-x_1-\sqrt{x_1^2-4\xi}\bigg)
\leq &x_2& \leq \frac{1}{2}\bigg( 2-x_1+\sqrt{x_1^2-4\xi}\bigg).
\end{eqnarray}
\end{subequations}
%-------------------

We observe that $\mathcal{C}^{\mu\nu}$ is the only factor
that has the $\Omega^*$ dependence in Eq.~(\ref{dC}).
The angular integral can easily be done if we express the
momenta $p'$ and $\bar{p}'$ in terms of $p_1$ and the 
relative momentum $q'$ as
%-------------------
\begin{subequations}
\label{cc-mom}
\begin{eqnarray}
p'&=&\tfrac{1}{2}p_1+q',
\\
\bar{p}'&=&\tfrac{1}{2}p_1-q',
\end{eqnarray}
\end{subequations}
%-------------------
where $p_1=(m_{c\bar{c}},\bm{0})$ and $q'=(0,\bm{p}_c^*)$
in the $c\bar{c}$ rest frame. After integrating out the
solid angle $\Omega^*$ analytically, we get
%-------------------
\begin{equation}
\label{intJ}
\int d\Omega^* \mathcal{C}^{\mu\nu}=
\frac{8\pi}{m_{c\bar{c}}^2}
\left(1-\frac{4|\bm{p}_c^*|^2}{3m_{c\bar{c}}^2}\right)
\left( -g^{\mu\nu} + \frac{p_1^\mu p_1^\nu}{p_1^2}\right).
\end{equation}
%-------------------
Eq.~(\ref{intJ}) behaves like the massive
spin-1 tensor with momentum $p_1$ with $p_1^2=m_{c\bar{c}}^2$.
Substituting Eqs.~(\ref{dPhi4_1}) and (\ref{intJ}) into Eq.~(\ref{dC}),
setting $P^2=4m_b^2$ at leading order in $v_b$, we simplify the
differential short-distance coefficient for the QCD process:
%------------------
\begin{equation}
\label{dc1-qcd}%
dC_1^{(\textrm{QCD})}=
\frac{(N_c^2-1)(N_c^2-4)}{N_c^3}
\frac{d\xi}{\xi}
\sqrt{1-\frac{r}{\xi}} \left(1+\frac{r}{2\xi}\right)
\frac{\alpha_s^4}{36\pi}
\int dx_1dx_2 \,F(\xi,x_1,x_2).
\end{equation}
%------------------
Here, the ranges of $x_1$ and $x_2$ integrals are given in
Eq.~(\ref{ranges}) and the integrand $F(\xi,x_1,x_2)$ is
given by
%------------------
\begin{equation}
\label{F} 
F(\xi,x_1,x_2)= \sum_{n=0}^{4}\frac{f_n(x_1,x_2)\,\xi^n}
{(x_1-2 \xi)^2 x_2^2 x_3^2},
\end{equation}
%------------------
where the coefficients $f_n(x_1,x_2)$ are
%------------------
\begin{subequations}
\label{fn} 
\begin{eqnarray}
f_0(x_1,x_2)&=& x_1^2(x_1 - 1)^2 + x_2^2(x_2 - 1)^2 
              + x_3^2(x_3 -1)^2,
\\
f_1(x_1,x_2)&=&-8+20(x_2+x_3)-20(x_2+x_3)^2+12x_2x_3+8(x_2+x_3)^3
\nonumber\\
&& -14x_2x_3(x_2+x_3)+x_2^2x_3^2,
\\
f_2(x_1,x_2)&=&6 - 16(x_2 + x_3)+ 12(x_2 + x_3)^2 
+2x_2x_3(x_2+x_3-6),
\\
f_3(x_1,x_2)&=&-4 + 8(x_2 + x_3) + 2 x_2 x_3,
\\
f_4(x_1,x_2)&=&2.
\end{eqnarray}
\end{subequations}
%------------------
Note that we use $x_3=2-x_1-x_2$ in Eqs.~(\ref{F}) and (\ref{fn}). 
The QCD contribution to the $c\bar{c}$ invariant-mass distribution 
in the inclusive $\Upsilon(1S)$ decay is finally obtained by replacing
$dC_1^{(c)}$ in Eq.~(\ref{Gam-c}) with $dC_1^{\textrm{(QCD)}}$
in Eq.~(\ref{dc1-qcd}).

Our result can be compared with a previous result in
Ref.~\cite{Fritzsch:1978ey}. Neglecting the term proportional to
$|\bm{p}_c^*|^2/m_{c\bar{c}}^2$ in Eq.~(\ref{intJ}),
we reproduce the function $\rho(\xi)$ in Ref.~\cite{Fritzsch:1978ey} 
analytically, where $\rho(\xi)$ is defined by
%-------------------
\begin{equation}
\label{rho}
\rho(\xi)=
\frac{d\Gamma[\Upsilon(1S)\to c\bar{c}+X]/d\xi}
     {\Gamma[\Upsilon(1S)\to \textrm{light hadrons}]}.
\end{equation}
%-------------------
In Ref.~\cite{Fritzsch:1978ey} the authors used the order-$\alpha_s^3$
color-singlet contribution to $\Upsilon(1S)\to ggg$ for the
$\Gamma[\Upsilon(1S)\to \textrm{light hadrons}]$. Imposing these 
approximations, we get
%-------------------
\begin{equation}
\label{rho2}
\rho(\xi)=
\frac{dC_1^{(\textrm{QCD})}/d\xi}
     {F_1({}^3S_1)},
\end{equation}
%-------------------
where $F_1({}^3S_1)$ is the short-distance coefficient 
of $\Upsilon(1S)\to ggg$ at leading order in $\alpha_s$ and 
$v_b$~\cite{Mackenzie:1981sf,Bodwin:1994jh,Bodwin:2002hg}:
%-------------------
\begin{subequations}
\label{F3S1}
\begin{eqnarray}
\Gamma[\Upsilon(1S)\to ggg]
&=&F_1({}^3S_1)
\frac{\langle \mathcal{O}_1({}^3S_1)\rangle_{\Upsilon(1S)}}
     {m_b^2},
\\
F_1({}^3S_1)&=&
\frac{(N_c^2-1)(N_c^2-4)}{N_c^3}\frac{(\pi^2-9)}{18}
\alpha_s^3.
\end{eqnarray}
\end{subequations}
%-------------------

Another check on the formula (\ref{dc1-qcd}) can be done in comparison
with a previous result for the color-octet spin-triplet $c\bar{c}$
contribution to the inclusive $J/\psi$ production in $\Upsilon(1S)$ decay:
Eq.~(\ref{dc1-qcd}) is to be compared with Eq.~(20) of
Ref.~\cite{Cheung:1996mh}. The function $F(\xi,x_1,x_2)$ in
Eq.~(\ref{F}) is equivalent to Eq.~(21) of
Ref.~\cite{Cheung:1996mh} up to an overall factor. After considering
the differences in the phase space and the normalization for the
states, we reproduce the results in Ref.~\cite{Cheung:1996mh} at
the leading order in $\alpha_s$, $v_b$, and the charm-quark velocity
$v_c$ in the $J/\psi$ rest frame ($\bm{p}_c^*=0$).
%------------------------------------------------
\subsection{\label{sec:collinear}Massless-charm-quark limit}
%------------------------------------------------
As is discussed earlier in this section, the short-distance coefficient
$dC_1^{(\textrm{QCD})}$ (\ref{dc1-qcd})
for the QCD process is free of IR and collinear divergences
as long as the charm quark is massive. However, in the limit where the 
charm quark becomes massless, $m_c\to 0$, $dC_1^{(\textrm{QCD})}$ may
acquire collinear divergences. These singularities cancel only if we
include the charm-quark contributions to the loop corrections to the
gluon wave functions in the process $\Upsilon(1S)\to ggg$
\cite{Bodwin:2007zf}. In the remainder of this section, we check
if $dC_1^{(\textrm{QCD})}$ (\ref{dc1-qcd}) satisfies correct
collinear behavior in the massless-charm-quark limit.

As the first step of the check, we can take
the limit $\xi\to 0$ on the function $F(\xi, x_1,x_2)$ in Eq.~(\ref{F}).
We find that $F(0,x_1,x_2)=f_0(x_1,x_2)$ and this value is proportional
to the color-singlet short-distance coefficient
$F_1({}^3S_1)$ in Eq.~(\ref{F3S1}) for the decay $\Upsilon(1S)\to ggg$
at leading order in $\alpha_s$ and 
$v_b$~\cite{Mackenzie:1981sf,Bodwin:1994jh,Bodwin:2002hg}:
%------------------
\begin{equation}
\label{F1LO}%
F_1({}^3S_1)=
\frac{(N_c^2-1)(N_c^2-4)}{N_c^3}\frac{\alpha_s^3}{18}
\lim_{\xi\to 0}\int dx_1dx_2 \,F(0,x_1,x_2).
\end{equation}
%------------------

Next, we investigate the asymptotic behavior of Eq.~(\ref{dC})
in the limit $m_{c\bar{c}}\to 2m_c$ and $m_c\to 0$.
As $m_{c\bar{c}}\to 0$, $|\bm{p}_c^*|$ approaches 
$m_{c\bar{c}}/2$ and gauge invariance requires 
$p_1$ to be orthogonal to  $\mathcal{B}_{\mu\nu}$.
By making use of Eqs.~(\ref{dPhi4_1}), (\ref{intJ}), 
and (\ref{dC}), we get
%-------------------
\begin{equation}
\label{BC-limit}
\mathcal{B}_{\mu\nu}
\int d\Phi_2(P\to p'+\bar{p}') \mathcal{C}^{\mu\nu}
\to
\frac{\mathcal{B}_{\mu\nu}}{2\pi m_{c\bar{c}}^2}\frac{1}{3}
\left( -g^{\mu\nu}\right).
\end{equation}
%-------------------
Note that, in this limit, $-\mathcal{B}_{\mu\nu}g^{\mu\nu}$
becomes the squared amplitude for $b\bar{b}_1({}^3S_1)\to ggg$.
Substituting Eq.~(\ref{dPhi4_1}) into Eq.~(\ref{dC}),
and using the limiting value (\ref{BC-limit}), we find that
the short-distance coefficient $C_1^{(\textrm{QCD})}$ is
divergent logarithmically in the limit 
$m_{c\bar{c}}\to 2m_c$ and $m_c\to 0$.
%------------------
\begin{equation}
\label{dc1-div}
C_1^{(\textrm{QCD})}\to
\frac{\alpha_s}{2\pi}
\left[
\frac{1}{8N_c}
\left(\frac{d^{abc}}{4\sqrt{N_c}}\right)^2
\left(4\pi\alpha_s\right)^3
\int \frac{d\Phi_3}{3!}
\left(-\frac{1}{3}g^{\mu\nu}\mathcal{B}_{\mu\nu}\right)
\right]_{p_1^2=0}
\int_{(2m_c)^2}^{(2m_b)^2}
\frac{dm_{c\bar{c}}^2}{m_{c\bar{c}}^2}
\end{equation}
%------------------
where the factor $8N_c$ comes from the perturbative NRQCD matrix
element in Eq.~(\ref{me-pert-norm}), the second and the third
factors in the brackets are the color factor and the coupling
for the process $\Upsilon(1S)\to ggg$, respectively, 
$1/3!$ is the symmetry factor for the three gluons,
and the factor $1/3$ is for the average over the initial spin
states. 
The quantity inside the square brackets in Eq.~(\ref{dc1-div})
is independent of $m_{c\bar{c}}$ and is finite.
Simplifying the leading
divergent term in Eq.~(\ref{dc1-div}), we obtain the asymptotic form
of $C_1^{(\textrm{QCD})}$. Because the collinear divergence is
absent in the QED contribution, the collinear divergent contribution
in the QCD process is the same as that in the short-distance 
coefficient $C_1^{(c)}$:
%-------------------
\begin{equation}
\label{dc1-qcd-divf}
C_1^{(c)}
\to \frac{\alpha_s}{\pi}\,
F_1({}^3S_1)\log\frac{m_b}{m_c}.
\end{equation}
%-------------------
It is explicit in Eq.~(\ref{dc1-qcd-divf}) that the collinear divergent
contribution, which is of order $\alpha_s^4$, is proportional to the 
short-distance coefficient for $\Upsilon(1S)\to ggg$ at order 
$\alpha_s^3$.

The only order-$\alpha_s^4$ contributions to $\Upsilon(1S)\to$ light
hadrons that depend on $m_c$ except for the $q\bar{q}gg$ final state
are the virtual charm-quark loop corrections to the gluon wave functions
in $\Upsilon(1S)\to ggg$. The leading divergent term of the virtual
correction is
%-------------------
\begin{eqnarray}
\label{dc1-virtual}
C_1^{(c,\textrm{virtual)}} &\to&
-3i\Pi(0)F_1({}^3S_1) \nonumber\\
&=& \frac{\alpha_s}{\pi}F_1({}^3S_1)\log\frac{m_c}{\mu},
\end{eqnarray}
%-------------------
where $\Pi(0)$ is the virtual charm-quark loop contribution to the 
vacuum polarization for an on-shell gluon and $\mu$ is
the renormalization scale. We find that the collinear divergence
cancels in the sum of the right sides of Eqs.~(\ref{dc1-qcd-divf})
and (\ref{dc1-virtual}). Therefore,
$C_1^{(c)}+C_1^{(c,\textrm{virtual})}$, which is 
the complete $m_c$ dependent contributions to the hadronic decay of
the $\Upsilon(1S)$ at order $\alpha_s^4$, is free of collinear divergence.

%------------------------------------------------
\section{\label{sec:rate}Numerical Analysis}
%------------------------------------------------
In this section, we provide a phenomenological prediction
for the $c\bar{c}$ invariant-mass distribution in inclusive
$\Upsilon(1S)$ decay by making use of the NRQCD factorization
formula obtained in Sec.~\ref{sec:cc}. As shown in Eq.~(\ref{dc1}),
the short-distance coefficient $dC_1^{(c)}$ is the sum of
the QCD and the QED contributions in Eqs.~(\ref{dc1-qcd})
and (\ref{dc1-qed}), respectively. Substituting the $dC_1^{(c)}$
into the NRQCD factorization formula (\ref{Gam-c}), we obtain the
differential rate depending on the scaled invariant mass
$\xi$ defined in Eq.~(\ref{variables1b}). The resultant $c\bar{c}$
invariant-mass distribution is 
%-------------------
\begin{equation}
\label{dmcc}
\frac{d}{dm_{c\bar{c}}^2}\Gamma[\Upsilon(1S)\to c\bar{c}+X]
=\frac{1}{m^2_{\Upsilon(1S)}}
\frac{dC^{(c)}_1}{d\xi}
\frac{\langle \mathcal{O}_1({}^3S_1)\rangle_{\Upsilon(1S)}}{m_b^2},
\end{equation}
%-------------------
where we use $P^2=m^2_{\Upsilon(1S)}$ and $m_{\Upsilon(1S)}$ is
the mass of the $\Upsilon(1S)$. In our numerical analysis,
we use the same input parameters as those used in Ref.~\cite{Kang:2007uv},
where the momentum distribution of the charm quark produced in the
inclusive $\Upsilon(1S)$ decay is studied.

The short-distance coefficient $dC_1^{(c)}$
depends on the strong coupling $\alpha_s$ and the ratio $r$
defined in Eq.~(\ref{variables1a}). For the strong coupling,
we take the running coupling $\alpha_s(m_{\Upsilon(1S)}/2) = 0.215$.
As shown in Eq.~(\ref{ranges-r}), the threshold of the phase space
is determined by the ratio $r=m_c^2/m_b^2$. In order to make the
end points of the phase space fit to the physical ones, we use
$m_c=m_D$ and $m_b=m_{\Upsilon(1S)}/2$ in evaluating the ratio $r$, where
$m_D=1.87$~GeV is the average mass of the $D^0$ and $D^+$ 
and  $m_{\Upsilon(1S)}=9.46$~GeV~\cite{Yao:2006px}. Then the numerical
value for the ratio becomes $r=4m_D^2/m_{\Upsilon(1S)}^2\approx 0.1563$.
This choice of $m_c$ and $m_b$ for the ratio $r$ seems reasonable for
the open-charm production in the $\Upsilon(1S)$ decay. For the 
bottom-quark mass $m_b$ that appears in the NRQCD factorization
formulas (\ref{Gam-c}) and (\ref{dmcc}), we use the one-loop pole
mass $m_b=4.6\pm 0.1$ GeV. The numerical value for the long-distance
NRQCD matrix element in Eqs.~(\ref{Gam-c}) and (\ref{dmcc}) is quoted
from Ref.~\cite{Kang:2007uv}:
%------------------
\begin{equation}
\label{ME}%
%------------------
\langle \mathcal{O}_1({}^3S_1) \rangle_{\Upsilon(1S)} =%
 3.07^{+0.21}_{-0.19}~\textrm{GeV}^3.
%------------------
\end{equation}
%------------------
For more details of the determination of the NRQCD matrix
element in Eq.~(\ref{ME}), we refer the readers to
Refs.~\cite{Bodwin:2007fz,Bodwin:2007ga,Bodwin:2006dn,Chung:2008sm}.

%%%%%%%%%%%%%%%%%%%%%%
\begin{figure}[tb]
\includegraphics*[width=12cm,angle=0,clip=true]{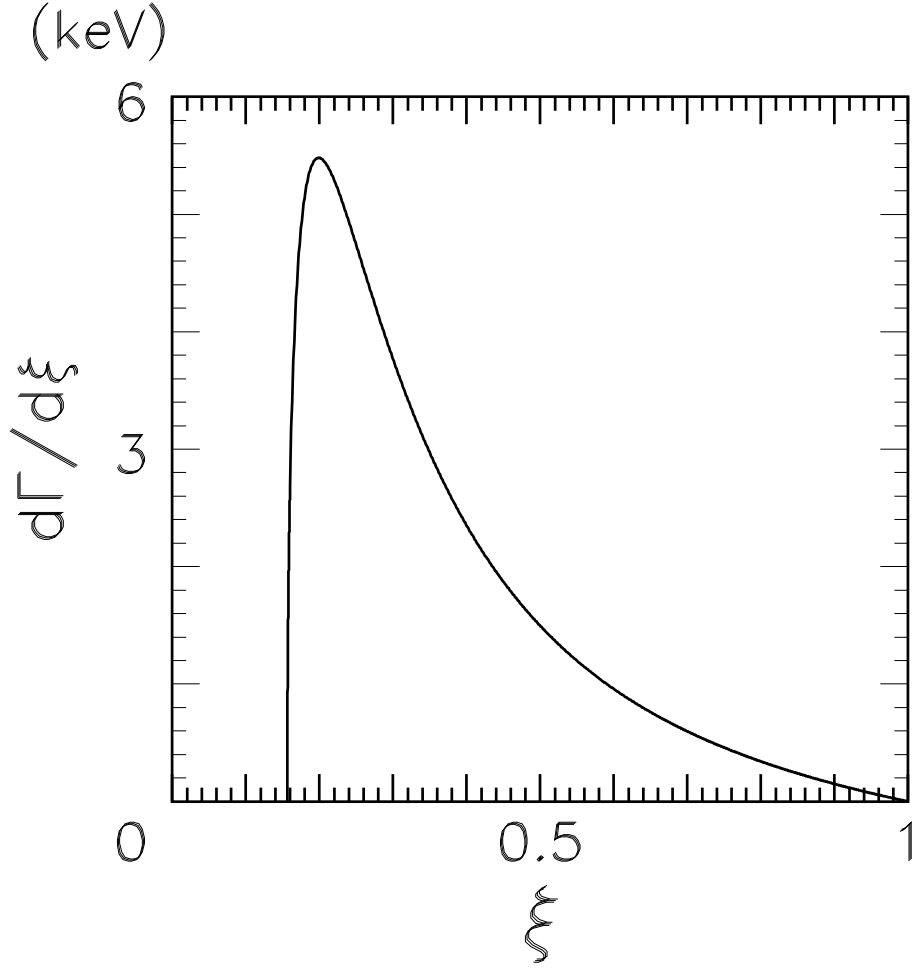}
\caption{\label{fig1}%
Invariant-mass distribution of the $c\bar{c}$ pair produced in the
inclusive $\Upsilon(1S)$ decay as a function of $\xi$, which is
defined in Eq.~(\ref{variables1b}). Only the contribution from the
QCD process $b{\bar b}({}^3S_1)\to c{\bar c}gg$ is shown. The
contribution from the QED process
$b{\bar b}({}^3S_1)\to\gamma^*\to c{\bar c}$,
which is proportional to $\delta(1-\xi)$, is not shown.}
\end{figure}
%%%%%%%%%%%%%%%%%%%%%%
The $c\bar{c}$ invariant-mass distribution from the QCD process
is obtained by substituting Eqs.~(\ref{dc1-qcd}) and (\ref{ME})
into Eqs.~(\ref{Gam-c}) and (\ref{dc1}), and integrating out
$x_1$ and $x_2$ over the ranges in Eq.~(\ref{ranges}).
We plot the QCD contribution to the invariant-mass
distribution $d\Gamma[\Upsilon(1S)\to c\bar{c}+X]/d\xi$
in Fig.~\ref{fig1} as a function
of the dimensionless variable $\xi$. Note that
$d\Gamma[\Upsilon(1S)\to c\bar{c}+X]/dm^2_{c\bar{c}}=%
m_{\Upsilon(1S)}^{-2}d\Gamma[\Upsilon(1S)\to c\bar{c}+X]/d\xi$.
The distribution in Fig.~\ref{fig1} has a sharp peak of
height $5.48\,$keV at $\xi=0.20$, which is just above the 
threshold at $\xi=r$.
The invariant-mass distribution of the $c\bar{c}$ from
the QED process is obtained in a similar way by using
Eq.~(\ref{dc1-qed}). As shown in Eq.~(\ref{dc1-qed}),
the QED contribution is proportional to the delta function
$\delta(1-\xi)$ so that the contribution is concentrated
at the kinematic end point $\xi=1$. Therefore, the contribution
of the QED process is well distinguished from the QCD contribution.

Integrating the invariant-mass distribution over the variable $\xi$,
we obtain the total production rate of the $c\bar{c}$ pair in the
inclusive $\Upsilon(1S)$ decay:
%------------------
\begin{subequations}
\label{total}%
\begin{eqnarray}
%------------------
\Gamma[\Upsilon(1S)\to c\bar{c}+X]&=&
\Gamma^{(\textrm{QCD})}+\Gamma^{(\textrm{QED})},
\\
\Gamma^{(\textrm{QCD})}&=& 1.44 \pm 0.36 ~ \textrm{keV},
\\
\Gamma^{(\textrm{QED})}&=& 2.60 \pm 0.65 ~ \textrm{keV},
%------------------
\end{eqnarray}
\end{subequations}
%------------------
where the theoretical uncertainties in Eq.~(\ref{total}) come from
the uncertainties of $m_b$, the NRQCD matrix element in Eq.~(\ref{ME}),
and uncalculated next-to-leading-order  relativistic and
QCD corrections, which we set to be 
10\,\% ($v_b^2\sim 0.1$) 
and
21.5\,\% ($\alpha_s= 0.215$) 
of the central value, respectively.
The QED contribution $\Gamma^{(\textrm{QED})}$ to $\Upsilon(1S)\to c\bar{c}+X$
agrees with that in Ref.~\cite{Kang:2007uv}. The QCD contribution
$\Gamma^{(\textrm{QCD})}$ differs from that in Ref.~\cite{Kang:2007uv} by
about $2$\,\% because we omit the contribution of the process
$b\bar{b}_1({}^3S_1)\to g^*g\gamma$ followed by $g^*\to c\bar{c}$
while this tiny contribution is included in Ref.~\cite{Kang:2007uv}.
As shown in Eq.~(\ref{total}), the QED contribution
$\Gamma^{(\textrm{QED})}$ occupies about 60\,\% of the total charm
production rate of the inclusive $\Upsilon(1S)$ decay. As is discussed in
Ref.~\cite{Kang:2007uv}, we expect that the QCD next-to-leading-order
corrections to the QED process
may modify the shape of the invariant-mass distribution.

Our result for $\Gamma^{(\textrm{QCD})}$ can be compared with
another previous result~\cite{Fritzsch:1978ey}.
In Ref.~\cite{Fritzsch:1978ey}, the ratio of the total charm
production rate in the inclusive $\Upsilon(1S)$ decay to
$\Gamma[\Upsilon(1S) \to \textrm{light hadrons}]$ is predicted 
to be $\Gamma[\Upsilon(1S)\to c{\bar c}+ X]/%
\Gamma[\Upsilon(1S) \to \textrm{light hadrons}]\approx 0.081 \,\alpha_s$.
This is greater by about $20$\,\% than our result,
\begin{equation}
\frac{\Gamma[\Upsilon(1S) \to c{\bar c}+ X]}
{\Gamma[\Upsilon(1S) \to \textrm{light hadrons}] }%
=0.065\,\alpha_s.
\end{equation}
As we discussed in Sec.~\ref{sec:cc},
the authors of Ref.~\cite{Fritzsch:1978ey} made the approximation
of neglecting $|\bm{p}_c^*|^2/m_{c\bar{c}}^2$ in computing the
function $\rho(\xi)$ in comparison with Eq.~(\ref{intJ}) of this
paper. The approximation leads to the overestimation
of the height of the peak in the invariant-mass distribution
and this is the reason for the discrepancy between the two results.
To check this point explicitly, we carry out the same calculation
with the approximations that were used in Ref.~\cite{Fritzsch:1978ey}.
The calculation shows that the height of the peak increases by 
about 8\,\% from $5.48\,$keV to $5.93\,$keV at the same horizontal
position at $\xi=0.20$. The difference increases as $\xi$ increases
ranging up to $30\,\%$. As a result, the total rate without
approximation is smaller by about $20\,\%$.
%------------------------------------------------
\section{\label{summary}Summary}
%------------------------------------------------
We have calculated the invariant-mass distribution of the $c\bar{c}$
pair produced in the inclusive decay of the $\Upsilon(1S)$ based
on the color-singlet mechanism of the NRQCD factorization formalism
at leading order in the bottom-quark velocity $v_b$ in the meson
rest frame. As the short-distance processes, we considered
the QCD process $b\bar{b}_1({}^3S_1)\to g^*gg$ followed by
$g^*\to c\bar{c}$ at leading order in $\alpha_s$ and the QED process
$b\bar{b}_1({}^3S_1)\to \gamma^*\to c\bar{c}$ at leading order
in $\alpha$ and $\alpha_s$.

The QCD contribution to the $c\bar{c}$ invariant-mass distribution
has a sharp peak just above the threshold, and that of the QED
process is concentrated at the maximally allowed kinematic end point.
In comparison with a previous analysis on the QCD process,
our prediction for the peak of the QCD contribution is lower than
that in Ref.~\cite{Fritzsch:1978ey} and the total production
rate of the $c\bar{c}$ pair in the inclusive $\Upsilon(1S)$ is smaller
than that in Ref.~\cite{Fritzsch:1978ey} by about 20\,\%. The main
reason for the discrepancy is that, in Ref.~\cite{Fritzsch:1978ey},
the authors made an approximation of neglecting
$|\bm{p}_c^*|^2/m_{c\bar{c}}^2$ while we keep the full expression
in Eq.~(\ref{intJ}) of this paper.

We also investigate the collinear divergences of the decay rate
in the massless charm-quark limit $m_c\to 0$. Although the decay
rate of leading order in $v_b$ is free of both IR and collinear
divergences if the charm quark is massive, the rate acquires
collinear divergences in the limit $m_c\to 0$. We have confirmed
that our analytic expression for the differential decay rate
reproduces the correct collinear behavior in this limit so that
the divergence exactly cancels that of the charm-quark loop 
corrections to the gluon wave function for the $\Upsilon(1S)\to ggg$
process. The sum of the two contributions are the $m_c$ dependent
contribution to the inclusive $\Upsilon(1S)$ decay into light
hadrons at order $\alpha_s^4$.
%--------------------------------------------------------------------
\begin{acknowledgments}
The work of HSC was supported by the BK21 program.
The work of TK was supported by the Korea Research Foundation under
MOEHRD Basic Research Promotion grant KRF-2006-311-C00020.
The work of JL was supported by the Korea Science and Engineering
Foundation (KOSEF) funded by the Korea government (MEST)
under Grant No. R01-2008-000-10378-0.
\end{acknowledgments}

\end{document}